\documentclass[useAMS,usenatbib]{mn2e}
\usepackage{color}
\usepackage{graphicx}
\usepackage{rotating}
\usepackage{lscape}
\usepackage{amssymb,amsmath}
\usepackage{url}
\usepackage{float}

\title[PC and AC relations]
{Empirical Period-Color and Amplitude-Color Relations for Classical Cepheids and RR Lyrae Variables}
\author[Bhardwaj et al.]{Anupam Bhardwaj$^1$\thanks{E-mail:
anupam.bhardwajj@gmail.com}, Shashi M. Kanbur$^2$, Harinder P. Singh$^1$, Chow-Choong Ngeow$^3$ \\ 
1. Department of Physics \& Astrophysics, University of Delhi, Delhi 110007, India. \\
2. State University of New York, Oswego, NY, USA.\\
3. Graduate Institute of Astronomy, National Central University, Jhongli 32001, Taiwan \\
}
\begin{document}

\date{Received on ; Accepted on }

\pagerange{\pageref{firstpage}--\pageref{lastpage}} \pubyear{2014}

\maketitle

\label{firstpage}

\begin{abstract}

We analyze Galactic, Large Magellanic Cloud and Small Magellanic Cloud 
Cepheids and RR Lyrae variables in terms of period-color (PC) and amplitude-color 
(AC) diagrams at the phases of maximum and minimum light. 
We compiled Galactic Cepheids $V$- and $I$-band data from the literature. We make use of optical bands light 
curve data from OGLE-III survey for Cepheids and RR Lyrae variables in the Magellanic Clouds.
We apply the $F$-statistical test to check the significance of any variation in the slope of PC and AC 
relations for Cepheid variables. The PC relation at maximum light for Galactic Cepheids with periods 
longer than about 7 days is shallow and the corresponding AC relation is flat for the entire period range.
For the fundamental mode Cepheids in the Magellanic Clouds, we find significant breaks in 
the PC and AC relations at both maximum and minimum light for periods around 10 days. 
The PC relation at maximum light for the Magellanic Clouds is flat for Cepheids with periods greater than 10 days. 
First overtone Cepheids with periods less than 2.5 days have a shallow PC relation at minimum light.
For fundamental mode RR Lyraes, we confirm earlier work supporting a flat 
PC relation at minimum light and a significant relation between amplitude and
color at maximum light. We find that no such relations exist for first overtone RR Lyrae stars.
These findings are in agreement with
stellar photosphere/hydrogen ionization front interaction considerations. 
These nonlinearities can provide strong constraints for models of stellar pulsation and evolution.

\end{abstract}

\section{Introduction}

\citet{smk1} investigated the period-color (PC) and amplitude-color (AC) relations at the phase of 
maximum, mean and minimum light for fundamental mode Cepheids in our Galaxy, 
Large Magellanic Cloud (LMC) and Small Magellanic Cloud (SMC).
These observational characteristics of Cepheids were analyzed following the work of \citet{smk93}, 
where full amplitude non-linear hydrodynamical pulsation models were used to explain the spectral 
class of Cepheids at maximum and minimum light. \citet{smk93} applied the 
Stefan-Boltzmann law at maximum and minimum light to show that

\begin{equation}
\log T_{max} -\log T_{min} = \frac{1}{10}(V_{min} - V_{max}),
\label{eq:temp_amp}
\end{equation}

\noindent where $T_{max}$ and $T_{min}$ are defined to be the effective photospheric 
temperature at maximum and minimum light, respectively. If 
$T_{max}$ or the color is independent of, or more weakly dependent on the pulsation 
period, then equation (1) predicts that there is a relation 
between the amplitude and the temperature at minimum light, and hence 
with color at minimum light. This relation suggests that higher
amplitude stars have cooler temperatures and hence redder colors at minimum light.

Conversely, if $T_{min}$ is independent or less dependent on the pulsation period, then the 
opposite is true: higher amplitude stars are driven to hotter temperatures and hence bluer colors at
maximum light. Consequently a relation between the amplitude and the temperature at 
maximum light is expected.

In other words, if for some reason, 
the PC relation at maximum (or minimum) light is flat or shallow,
then there is an AC relation at minimum (or maximum) light,  
suggesting a correlation between the amplitude and the extinction corrected 
color \citep{smk1, smk2}. Further, if the PC relation at maximum and/or 
minimum light undergoes a significant change in slope 
at a given period, equation (1) suggests that the AC relation at minimum 
and/or maximum light should change somewhat at 
that period too. Galactic Cepheids have shallow PC relations at maximum
light and an AC relation at minimum light \citep{smk93}.
In contrast, fundamental mode RR Lyrae stars have a flat PC 
relation at minimum light and a significant relation between amplitude and color at maximum light \citep{smk05}.

These studies on PC and AC relations were extended to observe 
multiphase characteristics by \citet{smk4}. These observational characteristics 
were further studied and published in a series of
papers by \citet{smk1}, \citet{smk2}, \citet{smk3}, \citet{smk4}, 
\citet{smk5} and \citet{smk6} for fundamental mode Cepheids. 

One reason the color at a particular phase and period may be 
weakly dependent on stellar pulsation parameters, producing a flat or flatter
PC relation at that phase could be due to the interaction 
of the stellar photosphere and hydrogen ionization front 
\citep[HIF,][]{smk1}. During a stellar pulsation cycle, the stellar photosphere and HIF
move in and out of the mass distribution and are not necessarily co-moving.
When the stellar photosphere and HIF are "engaged" (or the stellar photosphere lies at the base of the HIF), the
temperature of the stellar photosphere is close to the temperature 
at which hydrogen ionizes and hence the color of the star is close to the
color appropriate for the temperature at which hydrogen ionizes. 
The large rise in opacity due to the ionization of hydrogen makes it much harder for the
stellar photosphere to be pushed further inside the mass 
distribution of the star. Put another way, the temperature needed to produce a given
fraction of hydrogen ionization sufficient to prevent the photosphere 
from moving deeper inside the mass distribution is somewhat independent of global 
stellar parameters at low densities and temperatures. If this engagement 
occurs at higher densities, then the Saha ionization equation
predicts that higher temperatures are needed to produce the same
fractional hydrogen ionization and hence the same strength opacity wall. 
If the HIF and stellar photosphere are not engaged, then the
temperature of the stellar photosphere will again be more 
dependent on global stellar parameters such as period. 

Consequently, if for a group of stars, the stellar photosphere 
and HIF are engaged at low densities at a particular phase,
this will then lead to a flat or flatter PC relation at that 
particular phase for this group
of stars. For example, this is the case
with Galactic fundamental mode Cepheids at maximum light or 
fundamental mode RR Lyraes at minimum light. As a fundamental mode Cepheid dims from 
maximum light, the HIF moves further back inside the star and 
the HIF and photosphere become disengaged. Thus the temperature of the
photosphere or the color of the star becomes more dependent on 
global stellar parameters and thus on the period. Therefore, the PC relation at minimum light 
for fundamental mode Galactic Cepheids is not flat. Very long 
period (periods greater than about 20 days) Cepheids have their HIF so far inside the 
mass distribution that the stellar photosphere and HIF are not 
engaged at any point during the pulsation cycle, even at maximum light.  

In the case of fundamental mode RR Lyrae stars, as these stars 
brighten from minimum light, the HIF is driven further out in the mass distribution
and so the stellar
photosphere is pushed even further up the HIF, in order to 
achieve a given optical depth, say two-thirds. This necessitates a higher temperature
that is more dependent on global stellar 
parameters and the stellar pulsation period.

The pulsation phase and period at which such HIF-stellar 
photosphere engagement/disengagement occurs varies with 
metallicity and input physics such as the mass-luminosity (ML)
relation. This is because in hydrostatic models, the HIF moves further out in the mass distribution as
the $L/M$ ratio decreases and/or the $T_{eff}$ increases. Thus more centrally concentrated stars such
as RR Lyraes have their HIF further out in the mass distribution than Cepheids.
Because the $L/M$ ratio and $T_{eff}$ determine the HIF's location,
modeling of such observed nonlinearities appropriately has the potential to place
strong constraints on stellar pulsation models using ML
relations mandated by stellar evolution calculations.

The motivation for this paper was to extend the work on PC and AC relations with a much 
larger number of fundamental mode and first overtone Cepheids and RR Lyrae variables 
in the Magellanic Clouds based on the third phase of 
Optical Gravitational Lensing Experiment (OGLE-III) survey.
We also significantly extend the number and nature of Galactic Cepheids
included in our work over previous results. In particular we show that 
the empirical PCAC relations
are reasonably consistent with the HIF and stellar photosphere 
interaction theory described above and in earlier papers. Further, we extend the analysis
to include first overtone Cepheids and RR Lyraes found in the OGLE-III survey. 
The results are consistent with the stellar photosphere/HIF interaction 
theory outlined here and with the properties of the Saha ionization equation 
that is commonly used in models of stellar structure, pulsation and evolution.

In Section~\ref{sec:lc_data}, we describe the Galactic Cepheid light 
curve data compiled from 
literature  \citep{berdi08} and photometric light curve data for 
both Cepheid and RR Lyrae variables in the Magellanic Clouds taken 
from the OGLE-III survey \citep{oglelmcceph, oglelmcrrlyr, oglesmcceph, oglesmcrrlyr}.
In Section~\ref{sec:method}, we describe the method to evaluate the colors at the 
phases of minimum and maximum light from the Fourier fit to light curves. 
We also discuss the extinction correction in colors for all stars in 
Galaxy and Magellanic Clouds, and the $F$-statistical test for nonlinearity of the PC and AC relations.
In Section~\ref{sec:analysis}, we discuss the variation of PC and AC 
relations for Cepheids and RR Lyraes separately in the LMC, SMC and the
Galaxy. We compare our results for fundamental mode and first overtone stars and 
relate the changes with the theoretical models as described in the literature. We further discuss the robustness of our results 
in this section. We summarize the results and conclusions from this study in Section~\ref{sec:discuss}.


\section{the data}
\label{sec:lc_data}

The light curve data for fundamental mode Galactic Cepheids in $V$- and $I$-band have been 
compiled from the catalogue of \citet{berdi08}. 
The photoelectric observations for $V$- and $I$-band in the 
Johnson and Cousin photometric
systems were carried out by Berdnikov and 
his collaborators in series over nearly two decades between 1986 to 2004 \citep{berdi87, berdi92,
berdi93, berdit95, berdi98, berdit01, berdit03, berdit04}. We make use of 
348 Galactic Cepheids, which are common in $V$- and $I$-band, in our analysis. 
This compilation is a significant improvement in terms of the number of Cepheids 
to that compiled in previous work \citep{smk1}. The periods for the Galactic 
Cepheids were compiled from the McMaster catalogue of Galactic Classical Cepheids \citep{fernie95}.

The photometric light curve data in optical $V$- and $I$-band for 
Cepheids and RR Lyrae variables in the LMC and SMC was 
extracted from the OGLE-III survey.
We make use of fundamental mode (FU) and first overtone (FO) Cepheids 
in LMC and SMC from this survey \citep{oglelmcceph, oglesmcceph}, which include:
1802 FU and 1223 FO Cepheids in the LMC; and 2602 
FU and 1629 FO Cepheids in the SMC as classified by OGLE-III 
survey that are used in our analysis. 
 We also used the light curve data in optical bands for 26 long period fundamental mode Cepheids from
OGLE-III Shallow Survey in the LMC \citep{ulac13}. This increases our sample to 1828
fundamental mode Cepheids in the LMC.

Similarly, we also make use of fundamental mode RR Lyrae 
(RRab) and first overtone RR Lyrae (RRc) variables 
in the LMC and SMC from this survey \citep{oglelmcrrlyr, oglesmcrrlyr}. 
There are 17334 RRab and 4871 RRc stars 
in the LMC, while there are 1917 RRab and 171 RRc stars in the SMC. 
These stars have photometric light curve data in both $V$- and $I$-band.
We also used the time of initial epoch and the period as provided in OGLE-III database. 
We have summarized the data used in our analysis in Table~\ref{tab:data_table}.

Thus, one innovation in our paper is the large increase in data 
for Cepheids and RR Lyraes and the consideration of fundamental and first overtone
mode in both types of pulsating variables.

\begin{table}
\caption{The summary of the light curve data selected for the present analysis. These variables have light curve data in both $V$- and $I$-band.} 
\label{tab:data_table}
\scalebox{1.0}{
\begin{tabular}{ccccc}
\hline
\hline

  		&Variables	& Mode		& No. of stars 		& Reference   \\
\hline
\hline
Galaxy		&Cepheids	& FU		& 348			&B08\\
		
\hline
LMC		&Cepheids	& FU		& 1828			&S08,U13 \\
		&		& FO		& 1223			&S08 		\\

		&RR Lyraes	& FU		& 17334			&S09\\
		&		& FO		& 4871			& - 		\\
\hline
SMC		&Cepheids	& FU		& 2602			&S10a\\
		&		& FO		& 1629			& - 		\\

		&RR Lyraes	& FU		& 1917			&S10b\\
		&		& FO		& 171			& - 		\\

\hline
\hline
\end{tabular}}
{\footnotesize \textbf{Notes}: The references in the last column are, B08 \citep{berdi08}, S08 \citep{oglelmcceph}, S09 \citep{oglelmcrrlyr}, 
S10a \citep{oglesmcceph}, S10b \citep{oglesmcrrlyr}, U13 \citep{ulac13}.}
\end{table}

\section{The Method}
\label{sec:method}

\begin{figure*}
\begin{center}
\includegraphics[width=1.0\textwidth,keepaspectratio]{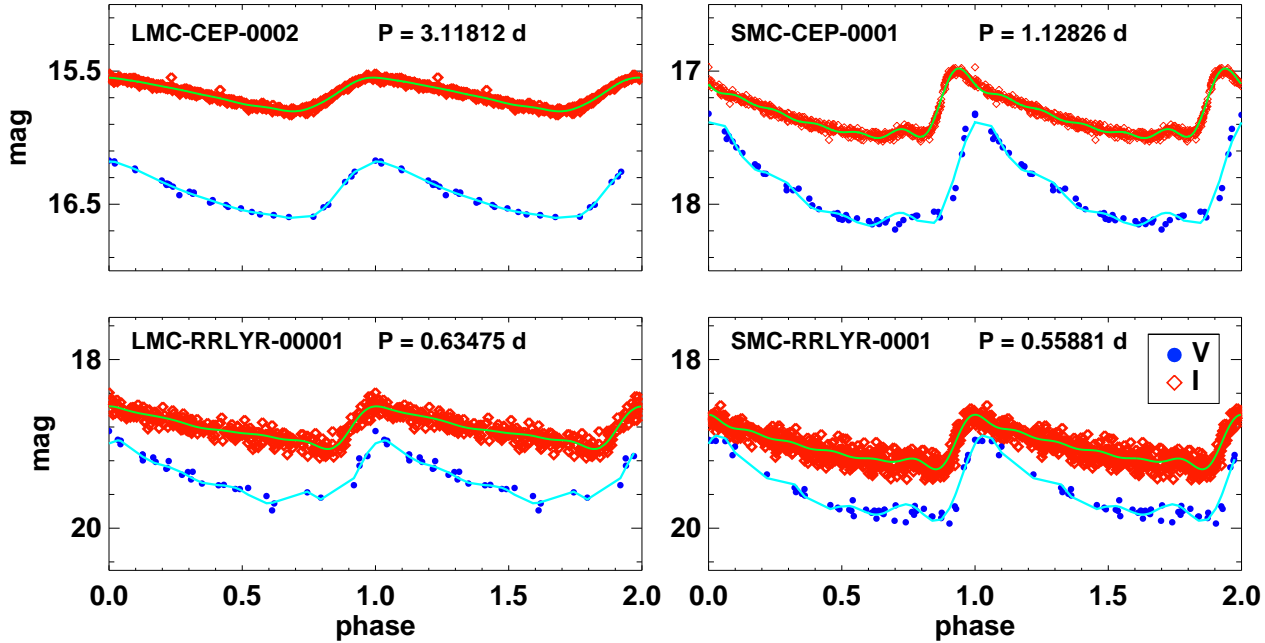}
\caption{Examples of Fourier fitted light curves of a Cepheid and a RR Lyrae variable each in the LMC and SMC from OGLE-III survey.}
\label{fig:lc_fit}
\end{center}
\end{figure*}

The Cepheid and RR Lyrae photometric light curve data in these three galaxies
were then fitted with a Fourier sine series \citep{slee81, deb10, deb14} with the following
form:

\begin{equation}
m = m_{0}+\sum_{k=1}^{N}a_{k} \sin(2 \pi k x + \phi_{k}),
\label{eq:foufit}
\end{equation}

\noindent where, 
\begin{equation*}
x = \frac{(t-t_{0})}{P} - \mathrm{int}\left(\frac{t-t_{0}}{P}\right).
\end{equation*}

\noindent Here, $t_{0}$ corresponds to the epoch of maximum brightness. This information is used to obtain 
a phased light curve that has maximum light at phase zero. Also, $m_{0}$ is the 
mean magnitude, $\emph{N}$ is the optimum order of the fit and $a_{k}$ and $\phi_{k}$ are the 
Fourier amplitude and phase coefficients. A low value for $\emph{N}$ 
can lead to a systematic deviation from the best estimate of calculated parameters whilst  
a high value for $\emph{N}$ may lead to over-fitting and numerical ringing \citep{peterson86}. So  
we varied the order of fit from 4 to 8 for each star and found the optimum order 
of fit corresponding to Bart's criterion \citep{barts82, deb09}. 
Examples of Fourier fitted light curves of a Cepheid and a RR Lyrae variable each in the LMC and the SMC 
are shown in Fig.~\ref{fig:lc_fit}. From the
optimum order fit, we obtain the V-band amplitude and colors at maximum and minimum 
light defined as :

\begin{gather*}
V_{amp} = V_{min} - V_{max} \\
(V-I)_{max} = V_{max} - I_{phmax} \\
(V-I)_{min} = V_{min} - I_{phmin}.
\end{gather*}

\noindent Here, $I_{phmax}$ and $I_{phmin}$ correspond to the $I$-band magnitude at the 
same phase as $V_{max}$ and $V_{min}$ \citep{smk1, smk3}.

\subsection{Extinction corrected color}

The extinction correction for LMC and SMC stars was carried out 
using the extinction maps given in \citet[][hereafter Haschke maps]{hasch11}.
We have used the RA/Dec position of the Cepheids and 
RR Lyraes from OGLE-III database to obtain the corresponding reddening from the Haschke maps.
The color excess $E(B - V)$ value for each Galactic Cepheid
were taken from the \citet{tammann03}. Finally, the colors at these two extremum phases 
have been corrected for extinction using $A_{V,I} = R_{V,I}E(B-V)$. The values
of R for Galactic data: $R_{V} = 3.17, R_{I} = 1.89$, are obtained from \citet{tammann03}. 
Similarly for the LMC and SMC,  
$R_{V} = 2.40, R_{I} = 1.41$ were used with color excess $E(V - I)$ values obtained from \citet{hasch11}.
We apply the same extinction values of $A_{V,I}$ to the colors at maximum and minimum 
light, since $A_{V,I}$ is independent of pulsation phase. 

\subsection{The $F$-statistical test}

Once we have the reddening
corrected colors, we plot them against $\log(P)$ and $V$-band amplitudes to find PC and AC relations.
To test for any possible breaks in the PC and AC relations, we use the $F$-test to 
determine the statistical significance of any variation in slope. We use the method described by 
\citet{smk1} to fit a single regression line and a double regression line 
separately to the PCAC relation under consideration. 
Therefore, under the null hypothesis, we can fit a regression line over the entire period range. 
As an alternative hypothesis, we can fit two regression
lines, one separately for stars having periods smaller/greater than the assumed 
break point. We will call
the former case (single regression line) to be the reduced model as described in \citet{smk1},
while the latter (two regression lines) is the full model. We write the first model as:

\begin{gather}
Y = a + bX ; ~~\mbox{Reduced model}. 
\end{gather}

\noindent This regression line is over the entire period range. To test a break at some point, we make use of the full model:

\begin{equation}
\begin{split}
Y = a_{S} + b_{S}X ; ~~\mbox{where P $<$ break point}\\
    ~~~~  = a_{L} + b_{L}X ; ~~\mbox{where P $\geq$ break point}.
\end{split}
\end{equation}

\noindent Here, $Y$ is the dependent variable, i.e. the color $(V-I)$ while $X$ is the independent variable, 
which is $\log(P)$ in case of PC relations and $V$-band amplitude in case of AC relations.

Our interest is to see whether the slope for short/long periods ($b_{S}$/$b_{L}$) varies significantly for the
break point under consideration. We can apply an appropriate F test statistic formulated as described in the following equation,

\begin{gather}
F = \frac{(RSS_{R} - RSS_{F})/[(n-2)-(n-4)]}{RSS_{F}/(n-4)},  
\end{gather}

\noindent where $RSS_{R}$, $RSS_{F}$ are the residual sums 
of squares in the single line regression (reduced model) and two
line regression (full model), respectively. Here, $n$ 
is the number of stars in the entire sample. 
The full model with two regression lines will have a smaller residual sum of squares. 
The null hypothesis that the single regression line is sufficient 
can be rejected if the the calculated
value of $F$ is greater than the critical value. The critical value 
is referred to the value of the $F$-distribution at 95$\%$
confidence level, i.e. $F_{C} = F_{2,n-4}$. 
Thus, a “large” value of $F$ provides evidence to support the 
rejection of the null hypothesis and hence the two regression line is a better fit to 
the data under consideration. The probability of 
the observed value of the $F$ statistic, $P(F)$, under the null hypothesis, 
gives the significance of this reduction in sum of squares. 
In all the plots for the PC and AC relations, we remove the 
$3\sigma$ outliers to have a robust regression analysis. 

In all subsequent tables, $a,b$ refer to the intercept and slope respectively and the subscripts $all,S, L$ refer to
the entire period range, short period range and long period range, respectively. The period separating short from long is mentioned in
the table title. 

\section{Analysis and Results}
\label{sec:analysis}

\subsection{Period-Color $\&$ Amplitude-Color relations for Cepheid variables}

\begin{figure}
\begin{center}
\includegraphics[width=0.5\textwidth,keepaspectratio]{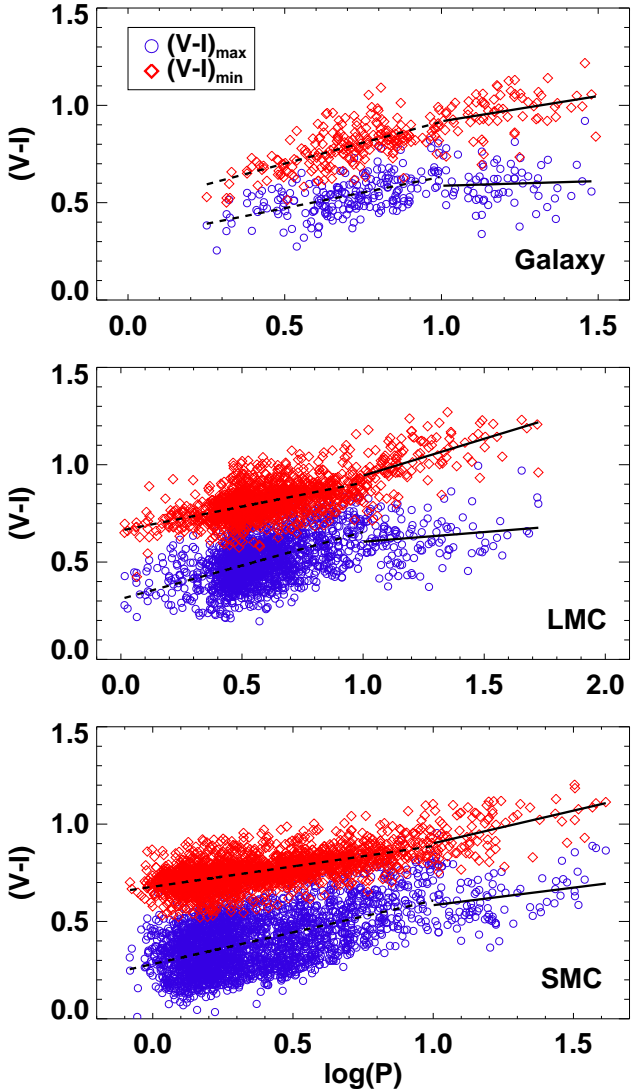}
\caption{ PC relations for fundamental mode Cepheids at maximum 
and minimum light for Galaxy, LMC and SMC. The dashed/solid lines represent the 
best fit to shorter/longer period Cepheids separated at 10 days.}
\label{fig:pc_fu_ceph.eps}
\end{center}
\end{figure}

\begin{figure}
\begin{center}
\includegraphics[width=0.5\textwidth,keepaspectratio]{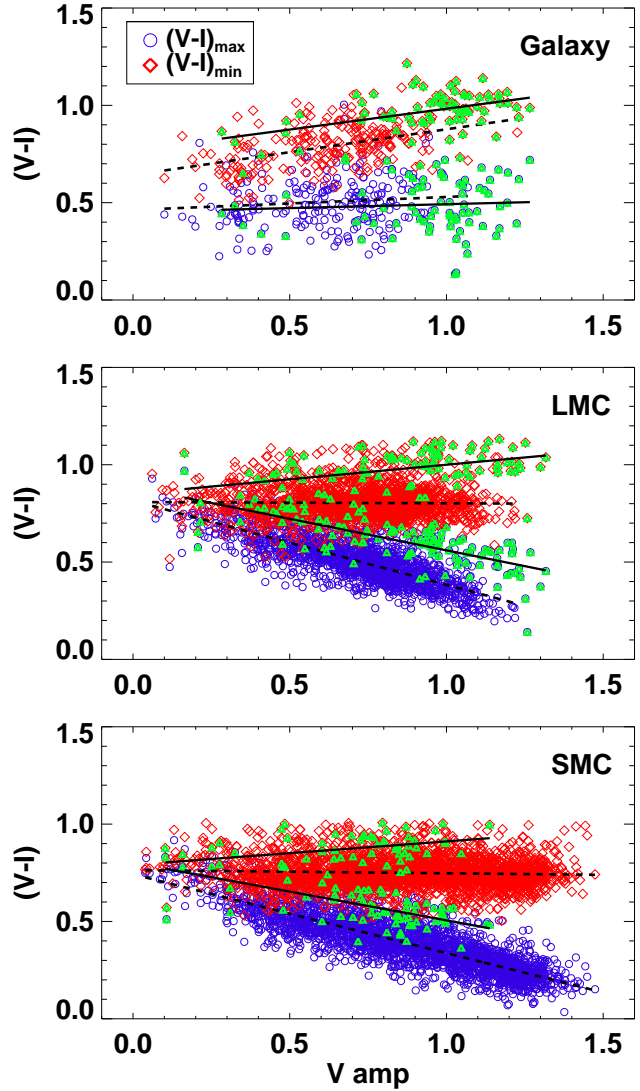}
\caption{ AC relations for fundamental mode Cepheids at maximum 
and minimum light for the Galaxy, LMC and SMC. Triangle represents the Cepheids having
periods greater than 10 days in both the plots at maximum and minimum light.
The dashed/solid lines represent the best fit to shorter/longer period Cepheids.}
\label{fig:ac_fu_ceph.eps}
\end{center}
\end{figure}

We describe the PC and AC relations for FU and FO Cepheids in the Galaxy, LMC $\&$ SMC in this subsection. 
The plots of PC and AC relations for FU Cepheids are given in Fig.~\ref{fig:pc_fu_ceph.eps}
and Fig.~\ref{fig:ac_fu_ceph.eps} respectively. 
Since, \citet{smk1} observed a break in the LMC PC 
relation for periods greater than 10 days, we fit the PC and AC relations to the entire sample as well as
the long and short period samples separated at 10 days.
The slope and intercepts for these PC and AC relations in the Galaxy,
LMC and SMC together with the results of $F$-test are provided in Table~\ref{table:pc_fu_ceph}. 

For Galactic FU Cepheids, the slope of the maximum light PC relation is significantly 
smaller for longer period ($P>10$~days) Cepheids than it is for their
shorter period counterparts. In fact this is also true for a break period of 7 days (or $\log(P)\sim 0.845$), 
as we see a flat PC relation at maximum light from a period of about 7 days instead of 10 days.
$F$-test results for a break period at 7 days
are presented in Table~\ref{table:fu_various_break}. Together with this we have a
statistically significant change in the slope of the AC relation at minimum light at both periods: 7 or 10 days. 
The AC relation at maximum light does not change significantly 
at 7 or 10 days. These findings are consistent with equation (1).

\begin{table*}
\begin{center}
\caption{Results of F test on PC and AC relations for fundamental 
mode Cepheids to determine possible nonlinearities at 10 days.}
\label{table:pc_fu_ceph}
\scalebox{1.0}{
\begin{tabular}{cccccccccc}
\hline
\hline
&Phase & b$_{(all)}$ & a$_{(all)}$ & b$_{S}$ & a$_{S}$ & b$_{L}$ & a$_{L}$  & F & P(F) \\
\hline
\hline
\multicolumn{10}{|c|}{Galaxy} \\
\hline
PC	&max&     0.177$\pm$0.021      &     0.406$\pm$0.018      &     0.322$\pm$0.034      &     0.311$\pm$0.024      &     0.048$\pm$0.107      &     0.539$\pm$0.128      &    11.527 &     0.000\\
	&min&     0.378$\pm$0.019      &     0.520$\pm$0.017      &     0.429$\pm$0.035      &     0.486$\pm$0.025      &     0.259$\pm$0.093      &     0.659$\pm$0.113      &     2.142 &     0.119\\
AC	&max&     0.024$\pm$0.033      &     0.483$\pm$0.025      &     0.067$\pm$0.049      &     0.463$\pm$0.032      &     0.040$\pm$0.075      &     0.452$\pm$0.072      &     1.075 &     0.343\\
	&min&     0.341$\pm$0.025      &     0.596$\pm$0.018      &     0.235$\pm$0.034      &     0.642$\pm$0.022      &     0.214$\pm$0.051      &     0.769$\pm$0.049      &    24.128 &     0.000\\
\hline
\multicolumn{10}{|c|}{LMC} \\
\hline
PC	&max&     0.253$\pm$0.011      &     0.359$\pm$0.007      &     0.343$\pm$0.015      &     0.311$\pm$0.009      &     0.099$\pm$0.056      &     0.505$\pm$0.069      &    32.022 &     0.000\\
	&min&     0.300$\pm$0.007      &     0.633$\pm$0.005      &     0.246$\pm$0.010      &     0.662$\pm$0.006      &     0.379$\pm$0.051      &     0.565$\pm$0.063      &    26.668 &     0.000\\

AC	&max&    -0.389$\pm$0.010      &     0.794$\pm$0.007      &    -0.432$\pm$0.009      &     0.815$\pm$0.007      &    -0.324$\pm$0.028      &     0.884$\pm$0.026      &   235.694 &     0.000\\
	&min&     0.035$\pm$0.010      &     0.788$\pm$0.008      &    -0.006$\pm$0.010      &     0.808$\pm$0.007      &     0.149$\pm$0.034      &     0.851$\pm$0.030      &   218.322 &     0.000\\

\hline
\multicolumn{10}{|c|}{SMC} \\
\hline
PC	&max&     0.302$\pm$0.008      &     0.288$\pm$0.004      &     0.326$\pm$0.011      &     0.281$\pm$0.004      &     0.180$\pm$0.073      &     0.404$\pm$0.088      &     7.381 &     0.001\\
	&min&     0.225$\pm$0.005      &     0.674$\pm$0.002      &     0.209$\pm$0.005      &     0.678$\pm$0.002      &     0.335$\pm$0.063      &     0.566$\pm$0.076      &    13.870 &     0.000\\
Ac	&max&    -0.408$\pm$0.005      &     0.749$\pm$0.005      &    -0.405$\pm$0.005      &     0.742$\pm$0.005      &    -0.296$\pm$0.041      &     0.801$\pm$0.033      &   128.116 &     0.000\\
	&min&    -0.019$\pm$0.006      &     0.769$\pm$0.005      &    -0.016$\pm$0.005      &     0.764$\pm$0.005      &     0.121$\pm$0.048      &     0.791$\pm$0.036      &    72.935 &     0.000\\

\hline
\end{tabular}}
\end{center}
\end{table*}

\begin{table*}
\begin{center}
\caption{Results of F test on PC and AC relations for fundamental 
mode Cepheids to check period breaks at 7 days for Galactic Cepheids and at 2.5 days for SMC Cepheids.}
\label{table:fu_various_break}
\scalebox{1.0}{
\begin{tabular}{cccccccccc}
\hline
\hline
&Phase & b$_{(all)}$ & a$_{(all)}$ & b$_{S}$ & a$_{S}$ & b$_{L}$ & a$_{L}$  & F & P(F) \\
\hline
\hline
\multicolumn{10}{|c|}{Galaxy} \\
\hline
PC 	&max&     0.177$\pm$0.021      &     0.406$\pm$0.018      &     0.286$\pm$0.052      &     0.333$\pm$0.033      &     0.042$\pm$0.046      &     0.552$\pm$0.049      &     7.539 &     0.001\\
	&min&     0.378$\pm$0.019      &     0.520$\pm$0.017      &     0.573$\pm$0.054      &     0.404$\pm$0.035      &     0.347$\pm$0.040      &     0.547$\pm$0.042      &     7.065 &     0.001\\
AC 	&max&     0.024$\pm$0.033      &     0.483$\pm$0.025      &     0.041$\pm$0.061      &     0.476$\pm$0.039      &     0.023$\pm$0.045      &     0.480$\pm$0.037      &     0.129 &     0.879\\
	&min&     0.341$\pm$0.025      &     0.596$\pm$0.018      &     0.265$\pm$0.038      &     0.600$\pm$0.025      &     0.257$\pm$0.030      &     0.708$\pm$0.026      &    37.111 &     0.000\\

\hline
\multicolumn{10}{|c|}{SMC} \\
\hline
PC	&max&     0.302$\pm$0.008      &     0.288$\pm$0.004      &     0.265$\pm$0.030      &     0.298$\pm$0.007      &     0.350$\pm$0.017      &     0.251$\pm$0.012      &     6.162 &     0.002\\
	&min&     0.225$\pm$0.005      &     0.674$\pm$0.002      &     0.176$\pm$0.016      &     0.685$\pm$0.003      &     0.263$\pm$0.009      &     0.645$\pm$0.007      &    15.078 &     0.000\\
AC	&max&    -0.408$\pm$0.005      &     0.749$\pm$0.005      &    -0.365$\pm$0.006      &     0.679$\pm$0.005      &    -0.419$\pm$0.007      &     0.816$\pm$0.006      &   590.648 &     0.000\\
	&min&    -0.019$\pm$0.006      &     0.769$\pm$0.005      &     0.015$\pm$0.006      &     0.709$\pm$0.006      &    -0.023$\pm$0.008      &     0.826$\pm$0.007      &   431.693 &     0.000\\

\hline
\end{tabular}}
\end{center}
\end{table*}

For FU Cepheids in the Magellanic Clouds at maximum light, we observe 
a significant change in the slope of the PC relation between
shorter ($P < 10$~days) and longer ($P > 10$~days) period Cepheids.
We find a significantly flatter PC relation at maximum light for FU Cepheids with periods greater 
than 10 days. In fact for LMC FU Cepheids, the longer period slope is consistent with zero to within the quoted errors. We note that
there are only two stars having period greater than 100 days, which were not considered in the best fit PC relations.
Further for SMC FU Cepheids, this shallow PC relation is observed for a small 
region of period range $1.0<\log(P)<1.3$. For stars with periods greater than 20 days, 
there was again a small but significant change in the slope of the PC relation. 
Now at minimum light, there is also a significant change in slope at 10 days for both the LMC and SMC.  
We also see that the short period Magellanic Clouds FU Cepheids PC relations at 
maximum and minimum light are almost parallel to each other.

For Galactic FU Cepheids, the AC relation at minimum light has a non-zero slope indicating that
higher amplitude FU Cepheids are driven to cooler temperatures. Hence the colors become redder at minimum
light because the PC relation at maximum light is flat
in accordance with equation (1). In the case of LMC and SMC FU Cepheids, the overall AC relations exhibit
a flat slope at minimum light but a positive slope at maximum light such that higher
amplitude stars are driven to hotter temperatures at maximum light. This is different to the Galactic AC relation at
minimum light (see Fig.~\ref{fig:ac_fu_ceph.eps}). These lower metallicity Cepheids in the Magellanic Clouds, 
 increase their amplitude by getting hotter at maximum light.
However, if we consider only the longer period ($P> 10$ days) FU Cepheids in the Magellanic Clouds, 
we see a significant change in the AC slopes at minimum light at $\log(P) \sim 1$ such that 
for $\log(P) > 1$, higher amplitude stars are driven to cooler temperatures and hence redder 
colors at minimum light, again in accordance with equation (1).
However the slopes in the AC relations for long period ($\log(P) > 1$) Cepheids are only marginally different from zero.

\begin{table*}
\begin{center}
\caption{Results for F test on PC and AC relations for first overtone Cepheids to check period breaks at 2.5 days.}
\label{table:fo_ceph}
\scalebox{1.0}{
\begin{tabular}{|c|c|c|c|c|c|c|c|c|c|}
\hline
\hline
&Phase & b$_{(all)}$ & a$_{(all)}$ & b$_{S}$ & a$_{S}$ & b$_{L}$ & a$_{L}$  & F & P(F) \\
\hline
\hline
\multicolumn{10}{|c|}{LMC} \\
\hline
PC&max&     0.105$\pm$0.008      &     0.484$\pm$0.003      &     0.099$\pm$0.011      &     0.485$\pm$0.003      &     0.334$\pm$0.042      &     0.362$\pm$0.022      &    10.384 &     0.000\\
&min&     0.085$\pm$0.008      &     0.639$\pm$0.003      &     0.086$\pm$0.010      &     0.640$\pm$0.003      &     0.235$\pm$0.046      &     0.557$\pm$0.024      &     5.216 &     0.006\\
AC&max&    -0.327$\pm$0.022      &     0.619$\pm$0.008      &    -0.373$\pm$0.026      &     0.627$\pm$0.009      &    -0.156$\pm$0.039      &     0.583$\pm$0.013      &    24.107 &     0.000\\
&min&     0.071$\pm$0.023      &     0.636$\pm$0.008      &     0.018$\pm$0.026      &     0.649$\pm$0.009      &     0.271$\pm$0.040      &     0.587$\pm$0.014      &    20.803 &     0.000\\
\hline
\multicolumn{10}{|c|}{SMC} \\
\hline
PC&max&     0.281$\pm$0.010      &     0.387$\pm$0.002      &     0.258$\pm$0.013      &     0.387$\pm$0.002      &     0.426$\pm$0.079      &     0.327$\pm$0.039      &     4.712 &     0.009\\
&min&     0.130$\pm$0.008      &     0.606$\pm$0.002      &     0.103$\pm$0.010      &     0.606$\pm$0.002      &     0.290$\pm$0.086      &     0.542$\pm$0.042      &     9.593 &     0.000\\
AC&max&    -0.490$\pm$0.011      &     0.649$\pm$0.005      &    -0.472$\pm$0.012      &     0.635$\pm$0.006      &    -0.198$\pm$0.041      &     0.600$\pm$0.015      &    85.904 &     0.000\\
&min&    -0.087$\pm$0.012      &     0.664$\pm$0.006      &    -0.069$\pm$0.012      &     0.649$\pm$0.006      &     0.196$\pm$0.042      &     0.616$\pm$0.015      &    74.455 &     0.000\\
\hline
\end{tabular}}
\end{center}
\end{table*}

\begin{figure*}
\begin{center}
\includegraphics[width=0.95\textwidth,keepaspectratio]{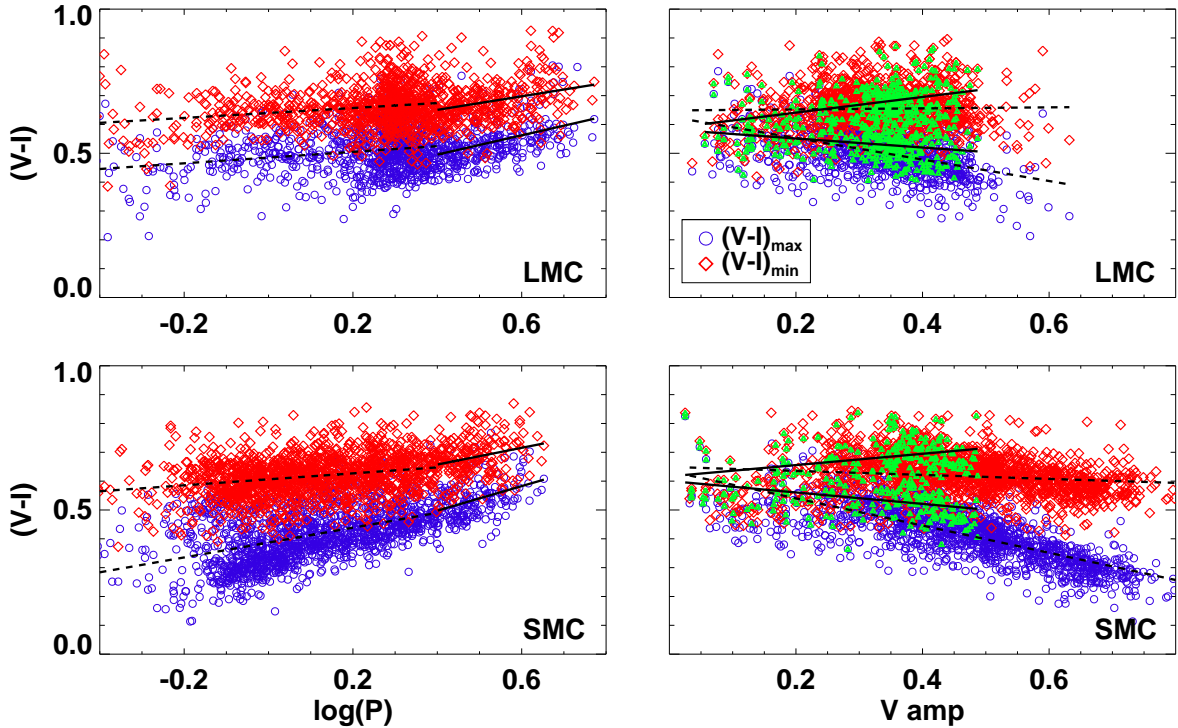}
\caption{PC and AC relations for LMC and SMC first overtone Cepheids 
at maximum and minimum light. The triangles represent the stars having
periods greater than 2.5 days in both the plots at maximum and minimum light.
The dashed/solid lines represent the best fit regression lines to shorter/longer period Cepheids.}
\label{fig:pc_ac_fo_ceph.eps}
\end{center}
\end{figure*}

It has been known that the period-luminosity relation for SMC FU Cepheids 
exhibits a break at a period of 2.5 days \citep[or $\log(P)=0.4$, see][]{bauer99}.
In addition, the plots of Fourier parameters against $\log(P)$ for SMC Cepheids show 
a change in progression around this period \citep{slee81}. We will comment on this further in the subsection describing 
robustness of our results. We present the results for PCAC breaks at a period of 2.5 days for 
SMC FU Cepheids in the second half of Table~\ref{table:fu_various_break}. 
These tests are highly significant at both maximum and minimum light. 
This analysis extends the earlier work 
of EROS collaboration \citep{bauer99} to the PC relation of OGLE-III $V$- and $I$-band data.

Fig.~\ref{fig:pc_ac_fo_ceph.eps} presents the PC and AC relations for LMC and SMC FO Cepheids. 
We also test the significance of any variation in slope of PCAC relations at 2.5 days for FO Cepheids in Magellanic Clouds.
The slopes and intercepts of these PCAC relations, together with the $F$-test results, are summarized in Table~\ref{table:fo_ceph}.
For Magellanic Clouds FO Cepheids, PCAC relations provide evidence of a break at 2.5 days at both maximum and minimum light.
We observed a flatter PC relation at minimum light 
and a greater slope at maximum light, for both LMC and SMC. The slope of the PC relation is almost flat for Cepheids having
periods less that 2.5 days.  
The AC relation at maximum light has a significant negative slope while we observe a flat AC relation at the 
phase of minimum light for both LMC and SMC. 

\begin{table*}
\begin{center}
\caption{Slope and intercepts for PC and AC relations for RRab and RRc stars at maximum and minimum light.}
\label{table:fu_fo_rrlyr}
\scalebox{1.1}{
\begin{tabular}{cccccccc}
\hline
\hline
   &  Phase  &   slope(RRab)   &  intercept(RRab)  &	$\sigma$(RRab) &   slope(RRc)   &  intercept(RRc) & $\sigma$(RRc)  \\
\hline
\hline
\multicolumn{8}{|c|}{LMC} \\
\hline
PC&max&     1.505$\pm$0.018      &     0.654$\pm$0.004      &     0.116&     0.770$\pm$0.032      &     0.638$\pm$0.015      &     0.084\\ 
&min&     0.093$\pm$0.019      &     0.716$\pm$0.005      &     0.116&     0.604$\pm$0.041      &     0.836$\pm$0.020      &     0.109\\
AC&max&    -0.361$\pm$0.003      &     0.592$\pm$0.002      &     0.091&    -0.089$\pm$0.014      &     0.312$\pm$0.007      &     0.091\\
&min&     0.049$\pm$0.003      &     0.651$\pm$0.003      &     0.114&      0.411$\pm$0.017      &     0.357$\pm$0.008      &     0.111\\
\hline
\multicolumn{8}{|c|}{SMC} \\
\hline
PC&max&     1.768$\pm$0.053      &     0.705$\pm$0.012      &     0.097&      0.536$\pm$0.228      &     0.529$\pm$0.100      &     0.079\\
&min&     0.055$\pm$0.058      &     0.725$\pm$0.013      &     0.099&      0.472$\pm$0.265      &     0.823$\pm$0.116      &     0.091\\
AC&max&    -0.370$\pm$0.007      &     0.594$\pm$0.006      &     0.074&    -0.312$\pm$0.078      &     0.450$\pm$0.039      &     0.074\\
&min&     0.067$\pm$0.010      &     0.660$\pm$0.008      &     0.098&      0.172$\pm$0.098      &     0.530$\pm$0.050      &     0.094\\
\hline
\end{tabular}}
\end{center}
\end{table*}

We also note that this behavior of a flatter slope at minimum light in the PC relation for FO Cepheids,
particularly those with periods less than 2.5 days,
is very similar to that observed for fundamental mode RR Lyrae stars as described in the next subsection.
We also comment on the selection of 2.5 days as break period in the subsection describing the robustness of our analysis.

\subsection{ Period-Color $\&$ Amplitude-Color relations for RR Lyrae variables }

\begin{figure*}
\begin{center}
\includegraphics[width=0.95\textwidth,keepaspectratio]{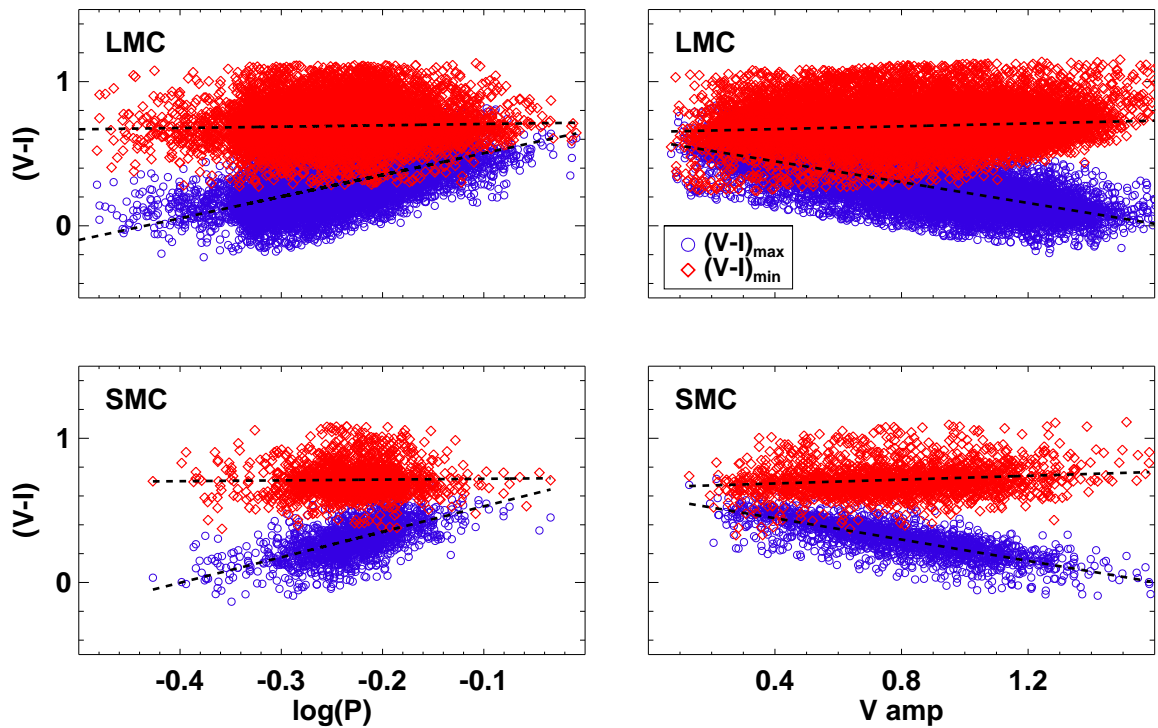}
\caption{ PC and AC relations for LMC and SMC RRab stars at 
maximum and minimum light. The dashed lines represent the best fit relations to the data.}
\label{fig:pc_ac_fu_rrlyr.eps}
\end{center}
\end{figure*}

\begin{figure*}
\begin{center}
\includegraphics[width=0.95\textwidth,keepaspectratio]{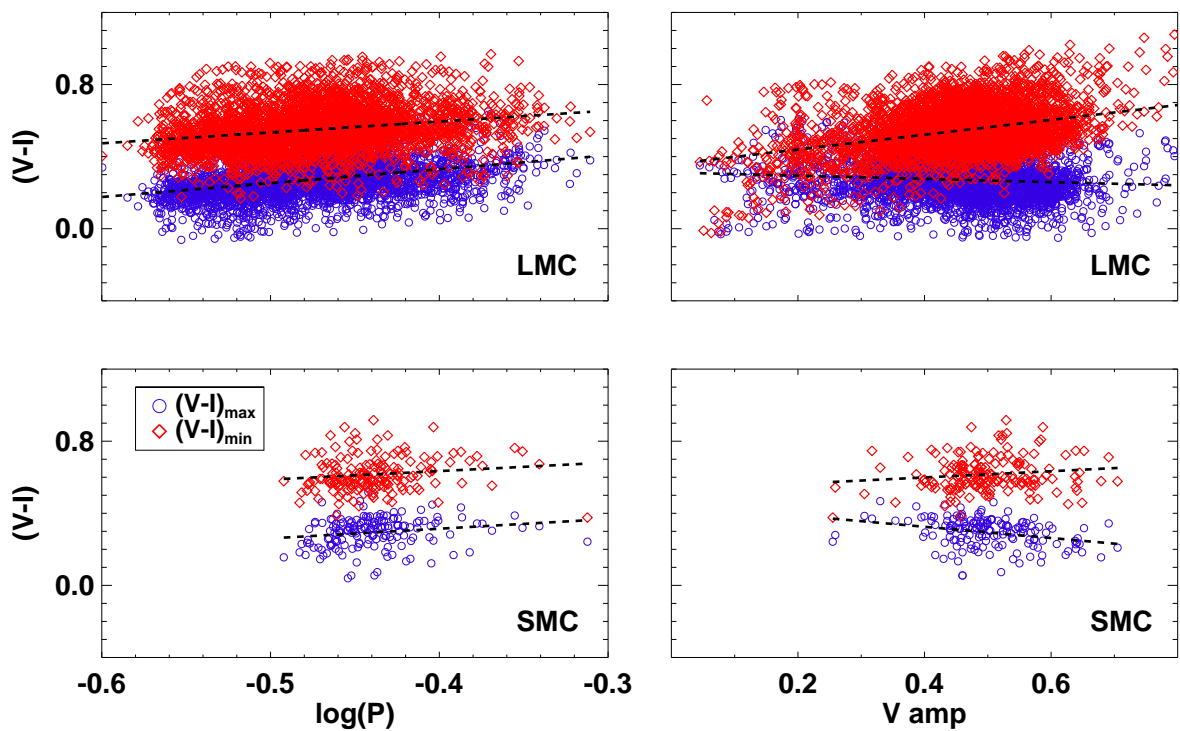}
\caption{ PC and AC relations for LMC and SMC RRc stars at 
maximum and minimum light. The dashed lines represent the best fit relations to the data.}
\label{fig:pc_ac_fo_rrlyr.eps}
\end{center}
\end{figure*}

A study of MACHO RRab stars in the LMC was carried out by \citet{smk05}. 
This suggested that for MACHO $(V-R)$ colors, PC and AC relations are
flat at minimum light (albeit with considerable scatter), but have a significant slope at maximum light.
We extend this analysis to RR Lyrae stars in the LMC and SMC based on the OGLE-III survey.
The plots of PC and AC relations for RRab and RRc stars in the Magellanic Clouds are presented in Fig.~\ref{fig:pc_ac_fu_rrlyr.eps} and 
Fig.~\ref{fig:pc_ac_fo_rrlyr.eps} respectively, while the results of fitting a 
linear regression to these PC and AC relations are summarized in Table~\ref{table:fu_fo_rrlyr}.
We find evidence to support a very shallow PC relation at minimum light 
in the $(V-I)$ color for both LMC and SMC RRab stars.
Similarly, the AC relation at minimum light is flat, while it has 
a significant negative slope at maximum light. Higher amplitude stars are
driven to bluer colors at maximum light, like FO Cepheids in the Magellanic Clouds, and in line with equation (1).

We note that though the PC relation at minimum light is very shallow,
there is considerable scatter. For the LMC/SMC the intercept of the PC relation at minimum light is
$0.716\pm0.005$/$0.725\pm0.013$, with a standard deviation on this relation of $0.116/0.099$, respectively. 
Since the PC relation is not exactly flat hence a mean value at minimum light is  $0.692\pm0.034$ and $0.711\pm0.043$
for LMC and SMC, respectively. We emphasize on this because there are a number of ways to define the minimum $(V-I)$ color. 
Our definition is quite specific and given in Section~\ref{sec:method} and is intended to capture the physics of the pulsation.

Previous work has suggested a common true intrinsic color for 
RRab stars at minimum light $\overline{(V - I)}_{min,0} = 0.57$ with an rms 
scatter of 0.025 mag \citep{slayden02} and 
$\overline{(V - I)}_{min,0} = 0.58\pm0.02$ mag \citep{gulden05}. These results use small samples of Galactic field RRab stars.
Our work involves 17334/1917 stars in the LMC/SMC, and in terms of scatter in the PC relation produces 
very similar results to the MACHO data \citep{smk05}.
However, the definition of the minimum light color in \citet{slayden02} and \citet{gulden05} is different to our analysis.
They have computed the colors at minimum light via an arithmetic mean of the observed $(V-I)$ data
values, having phases between 0.5 and 0.8. Hence, instead of a mean, if we consider a true minimum of $(V-I)$ data values, our
results are identical to those obtained using the definition $V_{min} - I_{phmin}$. 
In case of all the definitions, there is little difference in the minimum light color between the LMC and SMC.
Further, and perhaps more importantly, the scatter in the PC relation is quite large, indicating either
intrinsic scatter due to unknown causes, or the presence of significant extinction that is not accounted for.

We note that FO Cepheids in the Magellanic Clouds with $P<2.5$ days behave in a similar way to RRab 
stars in terms of PC and AC relations at maximum and minimum light. Both groups of stars
have a flat or relatively shallow PC relation at minimum light and a corresponding flat or shallow AC 
relation at minimum light. This flatness goes away for both PCAC relation at maximum light.
One possibility to explain this is that
FO Cepheids have hotter effective temperatures than FU Cepheids, hence the HIF in
a FO Cepheid lies much further out in the mass distribution \citep{kanbur95} causing the HIF and
photosphere to be engaged at minimum light in a similar way to RRab stars.
If this is confirmed by further theoretical calculations,
this will provide strong evidence that the HIF/stellar photosphere theory 
outlined in this and earlier papers is consistent with the observations.

The interesting thing that we observed for RRc stars is that the features we observe for RRab stars disappear for
both PC and AC relations in the LMC and SMC.
Recently \citet{layden13} have provided a PC relation for a small sample of field RRc stars as 
$(V-I)_{min,0} = 0.225 + 0.536P$, with an rms scatter of 0.073.
Our analysis suggests $(V-I)_{min} = 0.289 + 0.768P$ (with a standard deviation of 0.109) for LMC and
$(V-I)_{min} = 0.297 + 0.519P$ (with a standard deviation of  0.102) for SMC. 
Based on the PC and AC relations
presented in Fig.~\ref{fig:pc_ac_fo_rrlyr.eps} and Table~\ref{table:fu_fo_rrlyr}, 
the RRc PC relation at
minimum light is no longer flat but instead both PC and AC relations follow 
similar trajectories at maximum and minimum light.
Since RRc stars have in general hotter effective temperatures than RRab 
stars, RRc stars will have HIF's even further out in
the mass distribution as compared to RRab stars. The stellar 
photosphere will therefore be located quite high up the HIF. Further theoretical
work will be needed to investigate this possibility.

We also notice that the PC relations at maximum and minimum light for RRc stars in both the LMC and SMC are nearly parallel to
each other. This is also true for FO Cepheids in the LMC.

\subsection{Robustness}

\begin{figure*}
\begin{center}
\includegraphics[width=0.95\textwidth,keepaspectratio]{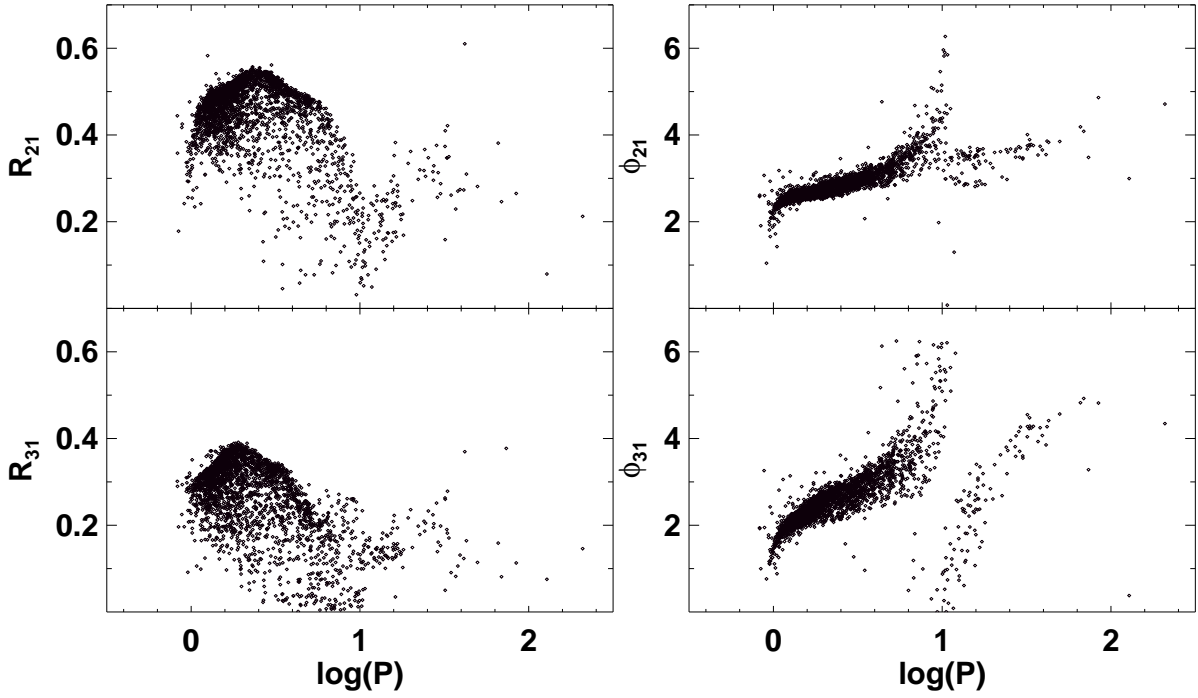}
\caption{Fourier parameters derived from Fourier decomposition technique for fundamental mode Cepheids in SMC using the $I$-band data.}
\label{fig:fou_params_fu.eps}
\end{center}
\end{figure*}

The assumptions behind the $F$-test are independence between observations, homoskedasticity and
normality of residuals.
Its clear that the observations are indeed independent of each other. We do see a trend in the residuals from our
straight line regressions but that is because the data are consistent with two lines. There is some reduction in scatter of the
residuals at periods close to 10 days but nothing to suggest that this would affect the distribution of the F statistic under the
null hypothesis. Also qq plots of the residuals do exhibit normality except perhaps at the extrema but our results hold true if these outliers
on the qq plots are removed \citep{smk1}. 

Our results do not change significantly even if we do not 
remove any outliers or if we do a $2\sigma$ clipping.
Further, moving the short period break from 2.5 days to 3.5 or 4.5 
days makes the various nonlinearities slowly go away for SMC FU Cepheids.
Similarly, for Magellanic Clouds FO Cepheids, the slope variations 
are no longer statistically significant if we shift the break point to
1 or 1.5 days. A similar situation holds for the longer period breaks at 10 days.

Further, our choice of applying a break period around 10 days is 
motivated, in part, by the way the structure of Cepheid light curves changes
with period around a period of 10 days \citep{slee81}. Sharp breaks are seen 
in the way the Fourier parameters change with period at 10 days. 

\begin{figure*}
\begin{center}
\includegraphics[width=0.95 \textwidth,keepaspectratio]{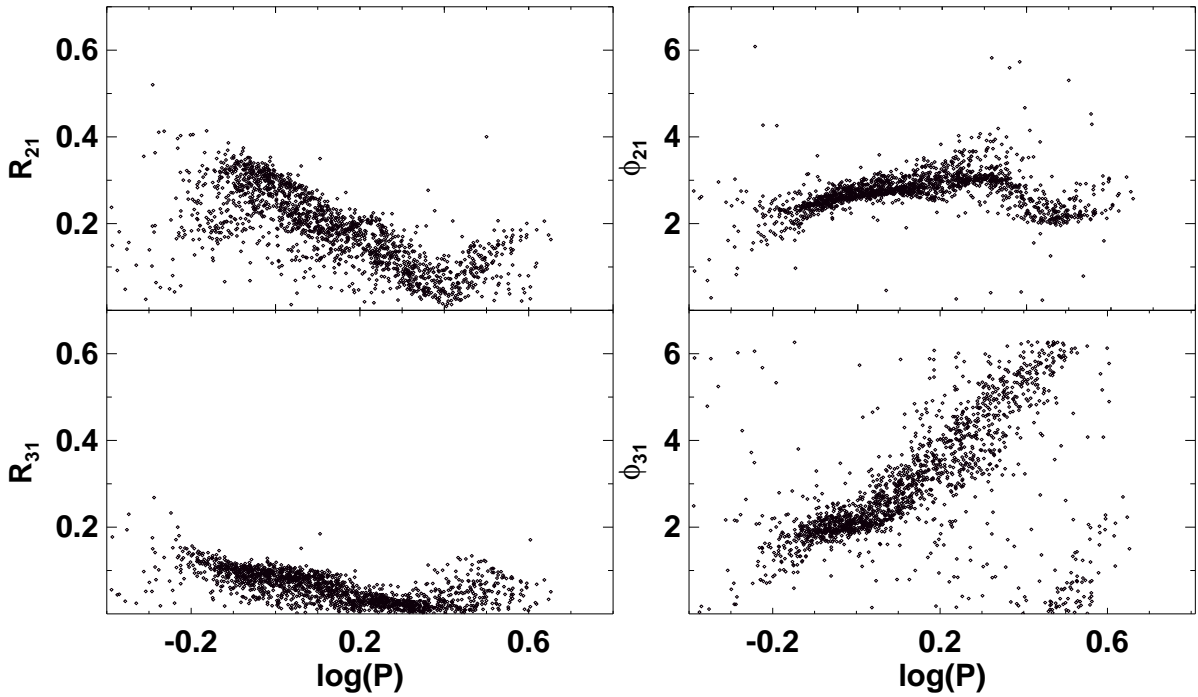}
\caption{Fourier parameters derived from Fourier decomposition technique for first overtone Cepheids in SMC using the $I$-band data.}
\label{fig:fou_params_fo.eps}
\end{center}
\end{figure*}

In a similar way, we can look at the plots of Fourier parameters ($R_{21}$, $R_{31}$, ${\phi}_{21}$ and ${\phi}_{31}$)
against $\log(P)$ for SMC Cepheids, based on the Fourier decomposition method \citep{slee81}.
These plots for SMC FU $\&$ FO Cepheids light curves
in the $I$-band, are displayed in Fig.~\ref{fig:fou_params_fu.eps} and Fig.~\ref{fig:fou_params_fo.eps}, respectively.
In the case of SMC FU Cepheids, the Fourier amplitude parameters presented in Fig.~\ref{fig:fou_params_fu.eps}, show a
maximum around 2.5 days ($\log(P)=0.4$). The Fourier phase parameters do not exhibit such patterns around
2.5 days but all parameters follow the change in progressions at 10 days.
Fig.~\ref{fig:fou_params_fo.eps} presents the 
way Fourier parameters change with periods around 2.5 days for FO SMC Cepheids. 
We see that there is a 
local minimum around $\log(P) = 0.4$, especially for the plot of $R_{21}$ against $\log(P)$.
Similar changes are observed in the Fourier parameters for LMC 
FO Cepheids. 

 Resonances in the normal mode spectra of Cepheids have provided a convincing explaination for the Cepheid bump Progression
($P_2/P_0 \approx 0.5, P_0 = 10$ days) and
support for other features in Cepheid light curves \citep{slee81, ander87}. Similarly, \citet{oglelmcceph} mentioned resonance
features for the first overtone pulsators at periods of about 0.35 and 3 days in OGLE-III data. The 
second feature has been interpreted as the signature of a 2:1 resonance between the first and fourth overtones
\citep{antone86}. Our ideas offer a detailed scenario through which
Period-Color relations and hence Period-Luminosity relations may change suddenly. These ideas are not mutually exclusive of
the resonane scenario. However, we note that our work can also be extended
to RR Lyraes. There are some indications of resonances in the RR Lyrae mode spectra, but nothing as yet, that seems
to stand out as significantly as the Cepheid $P_2/P_0$ resonance. The similarity in terms of PC relations at minimum light 
between FO Cepheids and FU RR Lyraes provides strong support for our ideas but again does not impinge on the effect of resonances
in the normal mode spectra.

\section{Discussion and Conclusions}\label{sec:discuss}

We have provided empirical evidence for the existence of a 
number of nonlinearities in PCAC relations at maximum and minimum light for 
Cepheids and RR Lyraes in the Galaxy and Magellanic Clouds. Specifically:

\begin{enumerate}
\item{The Galactic FU Cepheid PC relations at maximum 
and minimum light, together with the AC relation at minimum light,
exhibit a break at a period of 7 days.}

\item{The Galactic FU Cepheid PC relation at maximum light is 
flat for periods greater than 7 days, and is significantly different to the
PC relation at maximum light for shorter period counterparts.}

\item{At minimum light, neither the short nor long 
period Galactic FU Cepheids display a flat PC relation.}

\item{The Galactic AC relation at minimum light is such 
that higher amplitude stars are driven to cooler temperatures and
hence redder colors at minimum light.}

\item{For FU Cepheids in the Magellanic Clouds, there are PCAC 
breaks at 10 days at both maximum and minimum light. In both
cases the PC relation at maximum light for longer period Cepheids
is much shallower than for shorter period Cepheids. Similarly, 
in both Clouds the AC relation at minimum light goes from
having a slope consistent with zero for shorter period 
($P < 10$~days) Cepheids to having a positive slope for
longer period Cepheids. In addition the slope of the AC relation at maximum light
gets significantly lower across the 10 day transition.}

\item{In case of SMC FU Cepheids, we observe significant breaks in PC and AC relations 
around 2.5 days at maximum and minimum light.}

\item{In the case of FO Cepheids in Magellanic Clouds, 
there are PCAC breaks at both maximum and minimum light at a period around 2.5 days. More specifically,
for the LMC FO Cepheids, the PC relations at maximum and minimum light are very shallow for shorter period
($P<2.5$~days) Cepheids but much steeper for longer ($P>2.5$~days) period stars.
This is also true for the SMC PC relation at minimum light. 
However the SMC PC relation at maximum light is quite steep for shorter
period FO Cepheids ($P<2.5$~days) and becomes even steeper
for longer period FO Cepheids. For both LMC and SMC, the slope 
of the short period ($P<2.5$~days) AC relation at minimum light is very shallow and becomes
much steeper for
long period FO Cepheids. At maximum light the slope becomes 
less steep as one goes across the 2.5 days break period.}

\item{RRab stars in Magellanic Clouds have a flat PC relation 
at minimum light with a corresponding relation between amplitude and color at
maximum light such that higher amplitude stars have a hotter 
temperature and a bluer color at maximum light. Moreover, the 
flat PC relations at minimum light suggest that the the 
$(V-I)_{min}$ color in the LMC and SMC are consistent with each other to within the 
quoted errors.
Nevertheless, the dispersion of the PC relations around this 
minimum light color is quite significant.}

\item{The RRc stars in both the Magellanic Clouds do not exhibit 
such a flat PC relation at minimum light.}

\item{We note that the FO Cepheids in the Magellanic Clouds have a 
flat PC relation at minimum light just like RRab stars in the Magellanic Clouds.}

\end{enumerate}

In the following subsection we provide a possible theoretical scenario for some of these results.
The core of our theory relies on the interaction of the HIF and 
stellar photosphere and the properties of the Saha ionization
equation that is thought to determine ionization equilibrium in such stars.

\subsection{A possible theoretical interpretation}

The stellar photosphere and HIF are not generally co-moving during 
a pulsation cycle. The two can be engaged when the stellar
photosphere occurs at the base of the HIF and this engagement is 
sudden, i.e. they are either engaged or not engaged. When they are engaged
the temperature of the
stellar photosphere is the temperature at which hydrogen ionizes. 
Or rather the temperature of the stellar photosphere
is the temperature at which a large enough fraction of hydrogen 
is ionized to cause a substantial increase in opacity sufficient to
prevent the photosphere from moving further in the 
mass distribution of the star. Consequently, the color of the
star is the color corresponding to the temperature of 
the stellar photosphere for that particular star.
When this engagement occurs at low densities 
the PC relation is flat or shallow.
When this engagement occurs at high densities, a higher 
temperature is needed in order to achieve the
same level of hydrogen ionization and thus a large enough 
opacity to prevent the stellar photosphere from moving in any further in
the mass distribution. 
When the stellar photosphere and HIF are not engaged, the temperature of
the stellar photosphere is again dependent on period and global stellar parameters.

Changes in PC relations at maximum and minimum light 
can explain changes in AC relations due to equation (1). 
Thus (ii) occurs because for FU Cepheids, the HIF and 
stellar photosphere are 
engaged at low densities only at maximum light leading 
to a flat PC
relation at maximum light from a period of about 7 days 
onwards. Similarly, (iv) occurs because of equation (1) and the
fact that the PC relation at maximum light is flat. 

FU Cepheids in the Magellanic Clouds have a 
different ML relation that changes the 
relative location of the HIF and stellar photosphere 
such that the two are engaged at maximum
light, at low densities only for periods greater than 
10 days. Thus for Magellanic Clouds FU Cepheids with periods longer than 10 days, higher amplitude stars
are driven to cooler temperatures and hence redder colors at minimum light -- this explains (v).

Since RR Lyraes have hotter effective temperatures than Cepheids, the HIF lies 
further out in the mass distribution so that it is engaged with the
photosphere throughout the pulsation cycle. However, the engagement 
only occurs at low densities and temperatures at minimum light \citep{kanbur95}.

As the star brightens from minimum light, the HIF is 
pushed further out in the mass distribution. Thus the temperature of the stellar photosphere 
has to be hotter in order to get a greater fraction of 
hydrogen ionization in order to get a high enough opacity to get to a given optical depth.
Thus the color
at maximum light is not flat: higher temperatures are needed in order to get the
required hydrogen ionization fraction to achieve a high enough 
opacity to block further progress of the stellar
photosphere into the mass distribution. These required higher 
temperatures are dependent on period and global stellar parameters.
Thus the PC relation has a positive slope with a 
definite relation between amplitude and color at maximum light 
such that higher amplitude stars are driven to hotter temperatures and bluer colors
at maximum light because of equation (1). This explains (vii) and (viii). 

This behavior (a flat PC relation at minimum light) is not seen 
for RRc stars. Following the reasoning in \citet{kanbur95} we postulate that because 
overtone stars are generally hotter than fundamental mode stars, 
their HIF lies even further out in the mass distribution and the stellar photosphere
can get some way "up" the HIF before the high opacity prevents 
any further incursion. The temperature of the stellar photosphere would then be more
dependent on global stellar parameters such as the period. 
This suggests a possible explanation for (ix).  

We see from Table~\ref{table:fo_ceph} that FO Cepheids in 
the Magellanic Clouds with periods shorter than 2.5
days behave in a similar way to RRab stars. They have a flat PC
relation at minimum light. Correspondingly they have a well 
defined AC relation at maximum light and an AC relation with a slope consistent with zero at 
minimum light. This is again in accord with equation (1). 
FO Cepheids with periods shorter than 2.5 days
are hotter than FU Cepheids and their HIF lies further out in the mass distribution than
FU Cepheids. Hence in terms of the qualitative features 
of PCAC relations at maximum and minimum light, they behave just like RRab stars.
FO Cepheids with periods greater than 2.5 days have cool 
enough temperatures and higher $L/M$ ratios so that the stellar photosphere is disengaged from
the HIF and the temperature and hence color associated 
with the stellar photosphere would be more dependent on global stellar parameters, thus explaining (x).
 The theoretical concepts presented in this paper build on the theoretical calculations
carried out in \citet{smk93},\citet{kanbur95}, \citet{smk4}, 
\citet{smk5} and \citet{smk6}. In future work we plan a detailed theoretical study to 
determine if these ideas are consistent with the observations.

\subsection{Implications for the Cepheid Period-Luminosity relation}

The Cepheid Period-Luminosity (PL) relation is important for 
CMB independent estimates of Hubble's constant.
Since the PL relation is just the projection of the
Cepheid Period-Luminosity-Color (PLC) relation on the period 
and luminosity (or magnitude) planes, hence PC relations can affect PL relations through
this PLC relation. Both of the PL and PC relations at mean 
light are obtained by the numerical average of the corresponding PL and PC relations at
every phase during a pulsation period, therefore, changes in PC relations at particular phases can indeed
have effects on the mean light PC and PL relation. One way to understand PC
relations at mean light in greater depth is to study them 
as a function of phase, such as at maximum and minimum light. Thus this approach can also lead
to insights into possible nonlinearities in PC relations at mean light \citep[e.g., see][]{smk1}.
This is demonstrated here by the finding that the SMC FU Cepheid PCAC relations 
exhibit a highly significant break at a period of 2.5 days. 
This is consistent with the findings of \citet{bauer99} who found a break in the mean light
SMC PL relation for the FU Cepheids using EROS data.

We also note that our results pertain to the $V$- and $I$-bands. 
At longer wavelengths such effects become harder to measure because amplitudes
are smaller - temperature fluctuations lead to smaller effects 
in color as the wavelength increases due to the black body law. 
However any effects that happen at optical depths close to the photosphere could indeed
be affected by some of the physics discussed here - for example, colors at longer wavlengths that can be influenced
by opacity effects related to molecules.

\subsection{Further work}

We note that even though we have strong evidence for breaks in 
PCAC relations at a number of phases using the $F$-test and found these to be consistent
with a plausible theoretical scenario, more work needs to be done in order to be 
definitive about these results both from a theoretical and observational point of view.
More specifically, it is important to carry out a further range of statistical
tests, such as the testimator \citep{smk07test} and random walk \citep{koen07}, to be absolutely sure of
our observational results and stellar pulsation, structure and 
evolution modeling to further understand the nature of the HIF-stellar photosphere interaction.
We note that further investigation is needed before the SMC FU 
Cepheid and MC FO nonlinearities can be accommodated within
the framework of our stellar photosphere/HIF interaction theory. 
Thus one way to understand possible nonlinearities in Cepheid PL relations
is through the use of multiphase PC and PL relations as suggested here.

\section*{Acknowledgments}
\label{sec:ackno}
AB is thankful to the Council of 
Scientific and Industrial Research (CSIR), New Delhi for a Junior Research 
Scholarship (JRF). This work is supported by the grant provided by 
Indo-U.S. Science and Technology Forum under the Joint Center for Analysis of Variable Star Data.
SMK thanks SUNY Oswego for partial funding of research visits to the University of Delhi where some of this work was completed.
CCN thanks the funding from Ministry of Science and Technology (Taiwan) under the contract NSC101-2112-M-008-017-MY3.
The authors would like to thank Marcella Marconi, Ata Sarajedini, Sukanta Deb and Rachel Wagner-Kaiser for helpful discussions.  

\bibliographystyle{mn2e}
\bibliography{new_ac_pc}

\end{document}